\documentclass[10pt,conference]{IEEEtran}

\usepackage{xcolor}% Colored text

\usepackage{multirow} % For cells that comprise more than one line of a table
\usepackage{threeparttable}
  \usepackage[pdftex]{graphicx}
\usepackage[caption=false,font=footnotesize]{subfig}

\usepackage{stfloats}
\usepackage{amsmath}

\usepackage{amssymb}

\IEEEoverridecommandlockouts
\makeatletter
\def\ps@IEEEtitlepagestyle{
  	\def\@oddfoot{\Footer} % THIS PLACES THE FOOTER ON THE FIRST PAGE
	}
\let\old@ps@headings\ps@headings
\let\old@ps@IEEEtitlepagestyle\ps@IEEEtitlepagestyle
\def\confheader#1{%
  \def\ps@headings{%
    \old@ps@headings%
    \def\@oddhead{\strut\hfill#1\hfill\strut}%
    \def\@evenhead{\strut\hfill#1\hfill\strut}%
    \def\@oddfoot{\Footer} % THIS PLACES THE FOOTER ON ALL PAGES AFTER THE FIRST PAGE
  }%
  \def\ps@IEEEtitlepagestyle{%
    \old@ps@IEEEtitlepagestyle%
    \def\@oddhead{\strut\hfill#1\hfill\strut}%
    \def\@evenhead{\strut\hfill#1\hfill\strut}%
  }%
  \ps@headings%
}
\makeatother
\confheader{} % HEADER EXAMPLE 
\def\Footer{
   {\footnotesize
  \begin{minipage}{\textwidth}
  \centering \scriptsize{DISTRIBUTION STATEMENT A. Approved for public release: distribution is unlimited. Approval ID: $<$XXXX-2024-XXXX$>$.\\
  }  \end{minipage}
  }
}
\begin{document}
%
%
% PAPER TITLE
\title{Implementation of Field Programmable Gate Arrays (FPGAs) in Extremely Cold Environments for Space and Cryogenic Computing Applications}
%\title{Improvements to Discrete Low-Dropout Voltage Regulator Circuits for Extremely Cold Environments\\
 %\Large{Subtitle as needed}
%}
% AUTHOR NAMES AND AFFILIATIONS
% The separator \and provides for up to 3 columns
\author{\IEEEauthorblockN{Christopher Lewis }
\IEEEauthorblockA{Dept. of Elec. and Comp. Eng\\
Auburn University\\
Auburn, AL, USA\\
Email: cjl0027@auburn.edu}\\
\and
\IEEEauthorblockN{Drew Sellers}
\IEEEauthorblockA{Dept. of Elec. and Comp. Eng\\
Auburn University\\
Auburn, AL, USA\\
Email: jas0149@auburn.edu}
\and
\IEEEauthorblockN{Dr. Michael Hamilton}
\IEEEauthorblockA{Dept. of Elec. and Comp. Eng\\
Auburn University\\
Auburn, AL, USA\\
Email: mch0021@auburn.edu}\\

}
%Telephone: (800) 555--1212, Fax: (888) 555--1212}


\maketitle
% make the title


\begin{abstract}
The operation of CMOS Field Programmable Gate Arrays (FPGAs) at extremely cold environments as low as 4 K is demonstrated. Various FPGA and periphery hardware design techniques spanning from HDL design to improvements of peripheral circuitry such as discrete voltage regulators are displayed, and their respective performances are reported. While general operating conditions for voltage regulators are widened, FPGAs see a broader temperature range with improved jitter performance, reduced LUT delays, and enhanced transceiver performance at extremely low temperatures.

%if Keywords are used, end abstract text with \\

\end{abstract}
\renewcommand\IEEEkeywordsname{Keywords}

% \fi
%\IEEEpeerreviewmaketitle

\section{Introduction} 
With the increasing demand for exosphere-bound technologies in communication, large data/AI oriented imaging frameworks, and highly sensitive astronomical space observatories, reliable adaptation of commercial and complex technologies into environments of extreme temperature variations is paramount. By means of implementing commercial Field-Programmable Gate Arrays (FPGAs), a substantial reduction in cost, power, and development time can be realized while fulfilling these modern demands.

Inclusion of CMOS FPGA devices in cryogenic environments has been explored in recent works with a focus on RF multiplexing, as well as readout and control for cold computing technology \cite{ref1, ref2, ref3} given their current use to interface with these technologies from room temperature \cite{quench, digital, qubic}. These endeavors faced their own challenges, such as limiting the use of room temperature equipment to limit heat loading via interconnects for these cooling-power-limited cold stages. In response, recent designs have utilized low-dropout voltage (LDO) regulators constructed from discrete components \cite{ref2}. In practice, these designs face limited operational conditions that require the need for additional feed-through connections for different voltage domains, eliminating their benefits. Beyond their outlined applications as an enabler for quantum and neuromorphic computing, developments such as these are yet to be realized in similarly challenged environments such as satellite and astronomical space observatories that can seek to benefit from enhanced performances from these technologies and negating the requirement of complex thermal solutions.

In this paper, the operation of 28 nm MOSFET Xilinx ~Artix-7 FPGAs are demonstrated as well as results for the more powerful 16 nm FinFET Ultrascale+ counterparts that have seen limited exploration \cite{ref3}. Design requirements and considerations for the proper operation in these extremely cold environments will be discussed. Simultaneously, improvements to peripheral circuits, specifically voltage regulators, will be presented with their performance and in tandem with their required modes of operation.

\section{Hardware Description and Design}

\subsection{FPGA}

\begin{figure}[!t]
\centering
\includegraphics[width=2.5in]{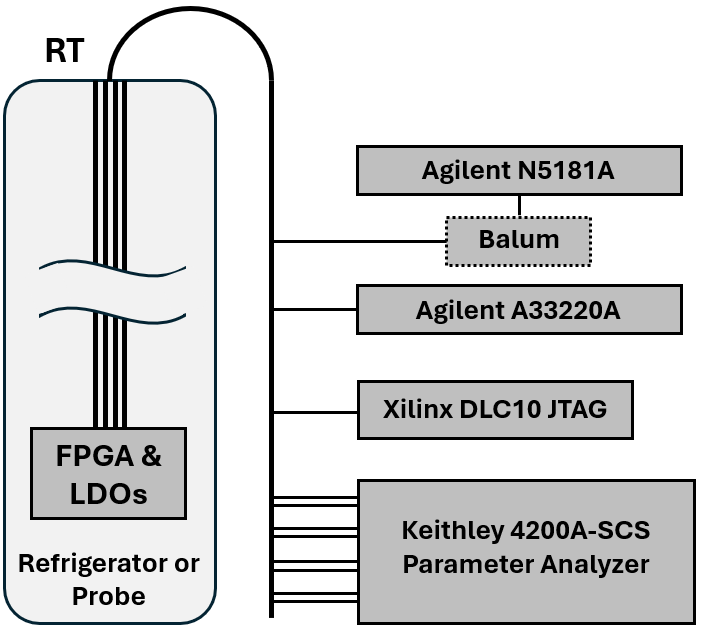}
\caption{Block diagram depicting the external clock, power, and programming sources for interfacing to remote FPGA and LDOs located in dilution refrigerators and/or dunk probes.}
\label{schematic}
\end{figure}

A standard 1.6 mm printed circuit board (PCB) of FR4 dielectric material is used to house these Xilinx FPGAs and the designed low-dropout (LDO) voltage regulators. For FPGA performance characterization, power is supplied externally via room temperature Source Measurement Units (SMUs) housed in a Keithley 4200A-SCS Parameter Analyzer with remote sensing enabled to negate Ohmic losses over lengthy cabling required for operation in refrigerators and dunk probes. Reference clock sources are also externally sourced via lab instrumentation (Agilent N5181A and Agilent A33220A) with an additional balum from Picosecond Pulse labs for differential clock sources due to previously observed degradation of crystal oscillators at lower temperatures. A simple schematic in Fig.\ref{schematic} provides a general visual overview of the testing infrastructure.

Two different families of Xilinx FPGA were selected for testing. The first was the power-optimized Artix-7 ~(XC7A35T-2CSG325I) due to its prevalence in previous works \cite{ref1, ref2, ref3} and its offering of gigabit transceivers. This device is constructed from the TSMC 28nm MOSFET HBM process node and features a single quad of GTP gigabit transceivers for transfer speeds up to 6 Gb/s that are connected to a sea of configurable logic. The second FPGA for testing was a Multi-Processor System on Chip (MPSoC) Zynq Ultrascale+ ~(XCZU1CG-1SBVA484I) for its low-cost entry into the Ultrascale+ family of devices. This device is of the TSMC 16 nm FinFET process node with it featuring a silicon-hardened processing side of the chip (PS) and a programmable logic section (PL) equivalent to that of a Xilinx Kintex FPGA in performance. The PS side of the architecture features a dual-core A53 ARM processor, a dual-core R5 ARM processor, several I/O circuits (UART, USB 3.0, I2C, gigabit Ethernet, etc.), and a DDR controller.

\begin{figure}[b]
\centering
\includegraphics[width=2in]{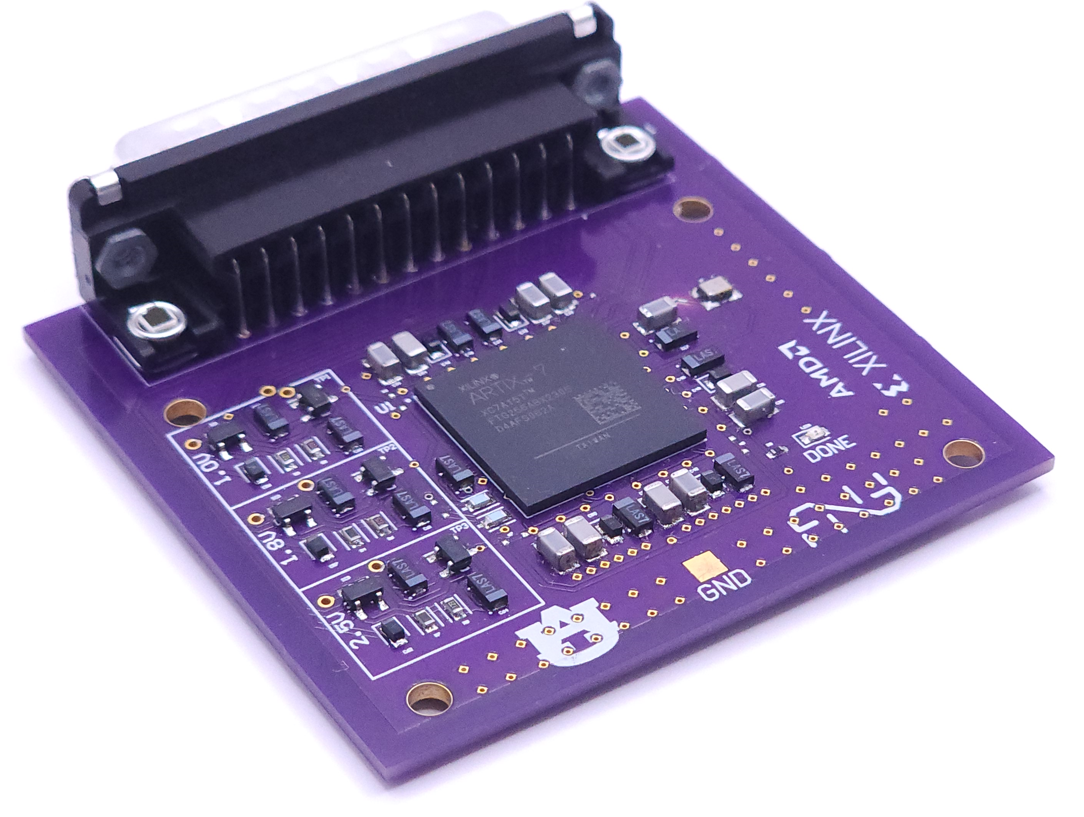}
\caption{Photo of the cryogenic assembly housing a Xilinx Artix-7 FPGA with its decoupling network and low dropout voltage regulators.}
\label{artix}
\end{figure}

\begin{figure}[h]
\centering
\includegraphics[width=2.5in]{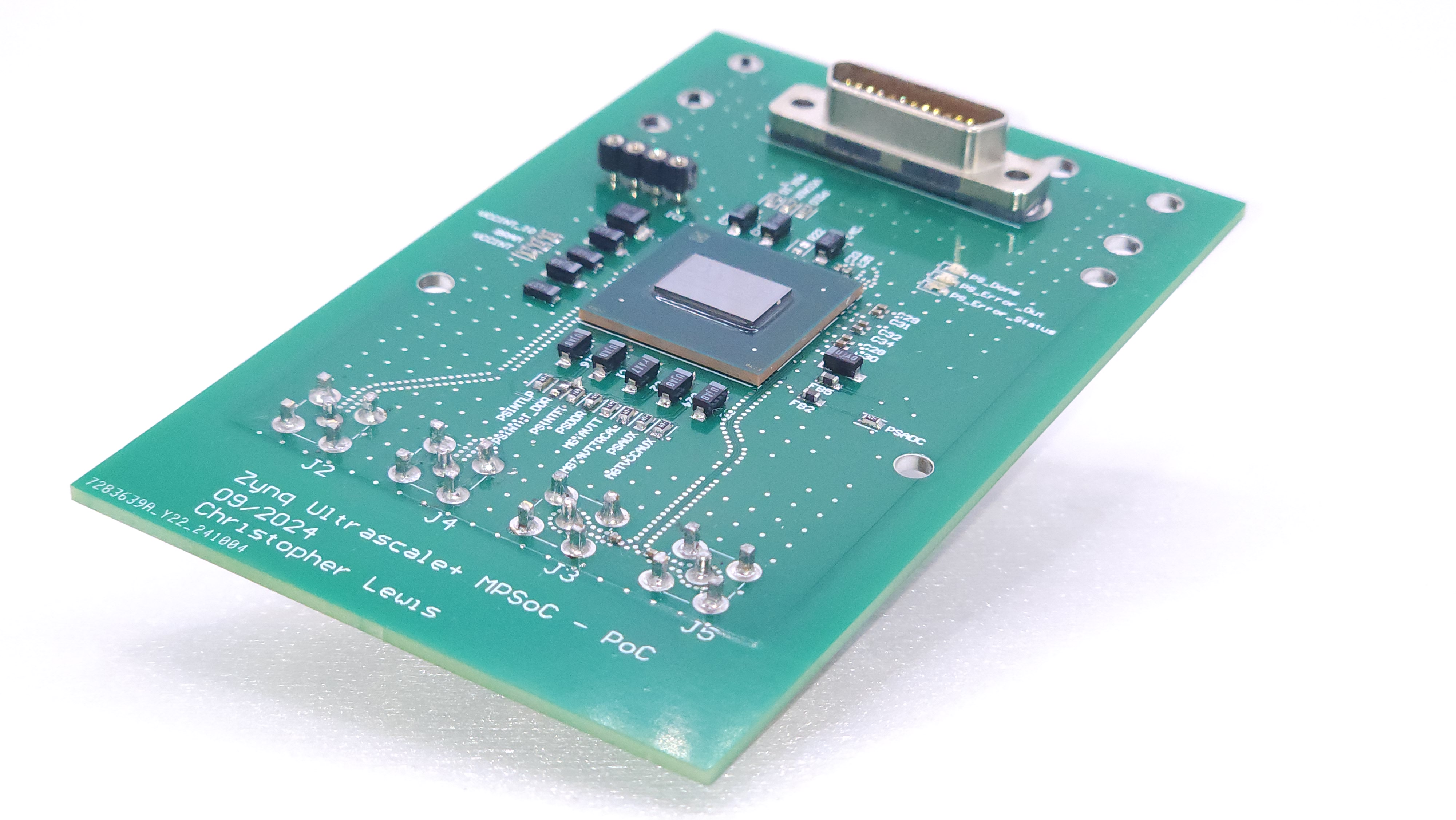}
\caption{Photo of the cryogenic assembly housing a Xilinx Zynq Ultrascale+ MPSoC with its decoupling network.}
\label{us+}
\end{figure}

Both families of FPGAs require considerable decoupling capacitors for power filtering and externally terminated resistors for reference bias. In conjunction with previous work, tantalum polymer capacitors are used for capacitances exceeding 1 uF, ~NP0-rated ceramic capacitors are used for lower capacitance values, and metal film resistors are used due to their proven effectiveness even at extremely low temperatures \cite{ref1}. In addition, no local flash memory was located on the board locally to the FPGAs, thus requiring remote JTAG programming after power-on. This methodology of design is consistent with previous work utilizing Artix-7 FPGAs at cryogenic temperatures, and is maintained with the Zynq Ultrascale+ FPGA given that its own implementation has not yet been documented in literature. These assemblies can be seen in Fig.~\ref{artix} and Fig.~\ref{us+}.

%Possibly include section of heat spreader and bubble migration hardware when and if it is done in time.

\subsection{Improvements to Voltage Regulators} 

To reduce room temperature voltage supply units and to simplify feedthrough connectivity, previous work proposed the use of discrete low dropout voltage regulators to replace ~commercially-available  solutions that display severally degraded performance in extremely cold environments \cite{ref3}. These regulators utilize general operational amplifiers (OpAmps) to compare and thus regulate the output in proportion to a known reference voltage. Operational conditions were found to be limited as they required initial power at room temperature before being cooled for proper regulation. 

\begin{figure}[!h]
\centering
\includegraphics[width=2.5in]{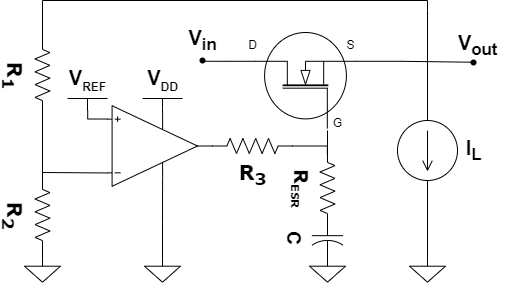}
\caption{Circuit diagram of the proposed cryogenic low dropout voltage regulator where $I_L$ is the output load current and $R_1$ and $R_2$ divide the output voltage to a common $V_{ref}$. The filter network for stability consists of $R_3$, $C$, and $R_{esr}$.}
\label{circuit}
\end{figure}

In this work, the use of a comparator IC (Microchip Technology MCP6541T-E/OT) is used in place of the OpAmp. Unlike the OpAmp, the comparator's output required filtering to prevent unwanted output ripple of the regulator. At 77 K, the regulator became far more susceptible to output rippage, which was resolved by adjustment of the equivalent series resistance (ESR) of the capacitor via a series resistor. The resulting circuit for the low dropout regulator is as seen in Fig.~\ref{circuit}.

%s shown in Fig.~\ref{opamp_sens}(b), the aforementioned decrease in input sensitivity disappears at lower temperatures (77 K). This leads to a scalable solution allowing multiple regulators to share a common V\textsc{in}, V\textsc{DD}, and V\textsc{ref}.

\section{Results}

\subsection{Commercial Evaluation Boards}
Implementation into extremely cold environments was first attempted using commercial evaluation boards featuring Xilinx FPGAs with AVNet's MicroZed System-On-Modules (SOMs). In these tests, failures owing to the board's on-board voltage regulators were seen early at 77 K within liquid nitrogen submersion within these regulated voltages deviating as far as 35\% from nominal. In this state, the FPGA was unable to communicate over JTAG for allowing remote programming and was unable boot from local flash memory. These regulators were then removed, and their respective voltage domains were driven by room temperature SMUs. Despite this, the SOMs' DRAM controllers were observed to never initialize properly, thus preventing FPGA startup. At the same time, flash memory became unresponsive through its interface, and the SOM's crystal oscillator was observed to deviate in frequency by over 20\% following previous works' observations \cite{ref2}. Overall, the use of evaluation boards for FPGA integration presents numerous sources of potential failures and serves as a demonstration of hardware design constraints that necessitate custom-made circuit boards that feature minimal peripheral circuitry.

\subsection{FPGA Operation and Performance}

An evaluation of the function of these FPGA technologies within cryogenic environments was performed by submersion into both liquid nitrogen and liquid helium, employing a dunk probe. This approach was used to provide adequate cooling power, thus ensuring uniform FPGA body temperatures across a broad range of power consumption levels. Power was supplied from room temperature via SMUs and local LDO regulators were bypassed to separate FPGA performance from regulator performance. As described previously, no local flash memory or crystal oscillators were located on the board, requiring JTAG remote programming and clock sources to also be sourced from room temperature.

Performance metrics for these FPGAs were collected including per-domain power consumption as well as combinational logic performance from the programmable logic (PL) section of the FPGA by reporting look-up table (LUT) delay in Table.~\ref{performance}. For comparison purposes, both device's transceivers were enabled due to its significant impact on power consumption. The power consumed by the processing system (PS) and programmable logic (PL) sections for the Zynq Ultrascale+ were summed together. LUT delay was measured via a 501-stage ring oscillator within the FPGA logic that is common practice for applications that measure circuit aging \cite{Xin}\cite{guin}. Due to place and route dictating RO performance, a percentage change is presented for comparison with existing work. Due to its level of power consumption, 4 K measurements were not performed for the Zynq Ultrascale+ device.

\begin{table}[ht]
\centering
\caption{Various performance characteristics across Xilinx's Artix-7 and Zynq Ultrascale+ FPGAs summarizing power consumption and combinational logic performance.}
\begin{tabular}{cccc|cc}
\hline
\multicolumn{1}{c|}{Device}           & \multicolumn{3}{c|}{\begin{tabular}[c]{@{}c@{}}Artix-7\\ (XC7A35T-2)\end{tabular}} & \multicolumn{2}{c}{\begin{tabular}[c]{@{}c@{}}Zynq Ultrascale+\\ (XCZU1CG-1)\end{tabular}} \\ \hline
\multicolumn{1}{c|}{Temperature (K)}  & 4                          & 77                        & 295                       & 77                                          & 295                                          \\ \hline
\multicolumn{1}{c|}{INT (mW)}         & 140                        & 138                       & 128                       & 738                                         & 774                                          \\
\multicolumn{1}{c|}{TT (mW)}          & 54                         & 54                        & 54                        & 118                                         & 132                                          \\
\multicolumn{1}{c|}{AUX/IO (mW)}      & 97                         & 62                        & 34                        & 136                                         & 101                                          \\ \hline
\multicolumn{1}{c|}{Total Power (mW)} & 291                        & 254                       & 216                       & 993                                         & 1007                                         \\ \hline
\multicolumn{1}{c|}{LUT Delay Change (\%)}   & -1.4\cite{performance}                        & -1.1                         &                          & -6.5                                         &                                           \\
\multicolumn{1}{c|}{LUT Spread Change(\%)}  & -16\cite{performance}                         & -2.0                        &                          & 6.1                                         &                                      
\end{tabular}
\label{performance}
\end{table}

As expected, the Zynq Ultrascale+ consumes five times the power of the Artix-7. The Artix-7 is a low-power device and contains only programmable logic. For both devices, the main power supply (INT) decreases with temperature while the transceiver's termination network (TT) remains consistent with the Artix-7 and slightly decreases with the Zynq Ultrascale+. This is expected for INT, as it consists mostly of dynamic power. For both devices, the mixture of voltage domains for auxiliary power and input/output interfaces, AUX/IO, is seen to increase with decreasing temperature. This increase is expected because auxiliary power consuming the majority out of the two domains and consists largely of static power.

Additional testing was also conducted for the modules within the PS-side of the FPGA for basic operation while at cryogenic temperatures. As can be seen in Table.~\ref{PS}, all tested modules were found to be operational with the DDR controller module being tested with CMOS LPDDR3-1600 commercial memory module which was located next to the FPGA device.

\begin{table}[h]
\centering

  \begin{threeparttable}

\caption{Basic Operation for all modules featured in the Processing system (PS) section of the Zynq Ultrascale+ FPGA at 77 K.}
\begin{tabular}{c|c}
\hline
PS Module                                                             & 77 K Operation \\ \hline
APU                                                                   & \checkmark             \\

RPU                                                                   & \checkmark               \\

\begin{tabular}[c]{@{}c@{}}IOU (GPIO, SPI, USB, etc.)\end{tabular} & \checkmark              \\
OCM                                                                   & \checkmark               \\
DDR Controller                                                       & \checkmark$^1$           \\
\begin{tabular}[c]{@{}c@{}}GTR (PCIe, SATA, DP)\end{tabular}        & \checkmark               \\
DMA (FPD, LPD)                                                        & \checkmark               \\
PL Interface Ports                                                    & \checkmark              
\end{tabular}
\raggedleft
\begin{tablenotes}
       \item \checkmark - Successful operation.
       \item $^1$ - Tested with LPDDR3-1600
       %\item $\setminus$ - Untested.
     \end{tablenotes}
\label{PS}
  \end{threeparttable}

\end{table}
%\footnotesize{$^a$ The smallest spatial unit is county, $^b$ more details in appendix A}\\

\subsection{Low-dropout Regulators}

\begin{figure}[!h]
\centering
\subfloat[]{\includegraphics[width=1.7in]{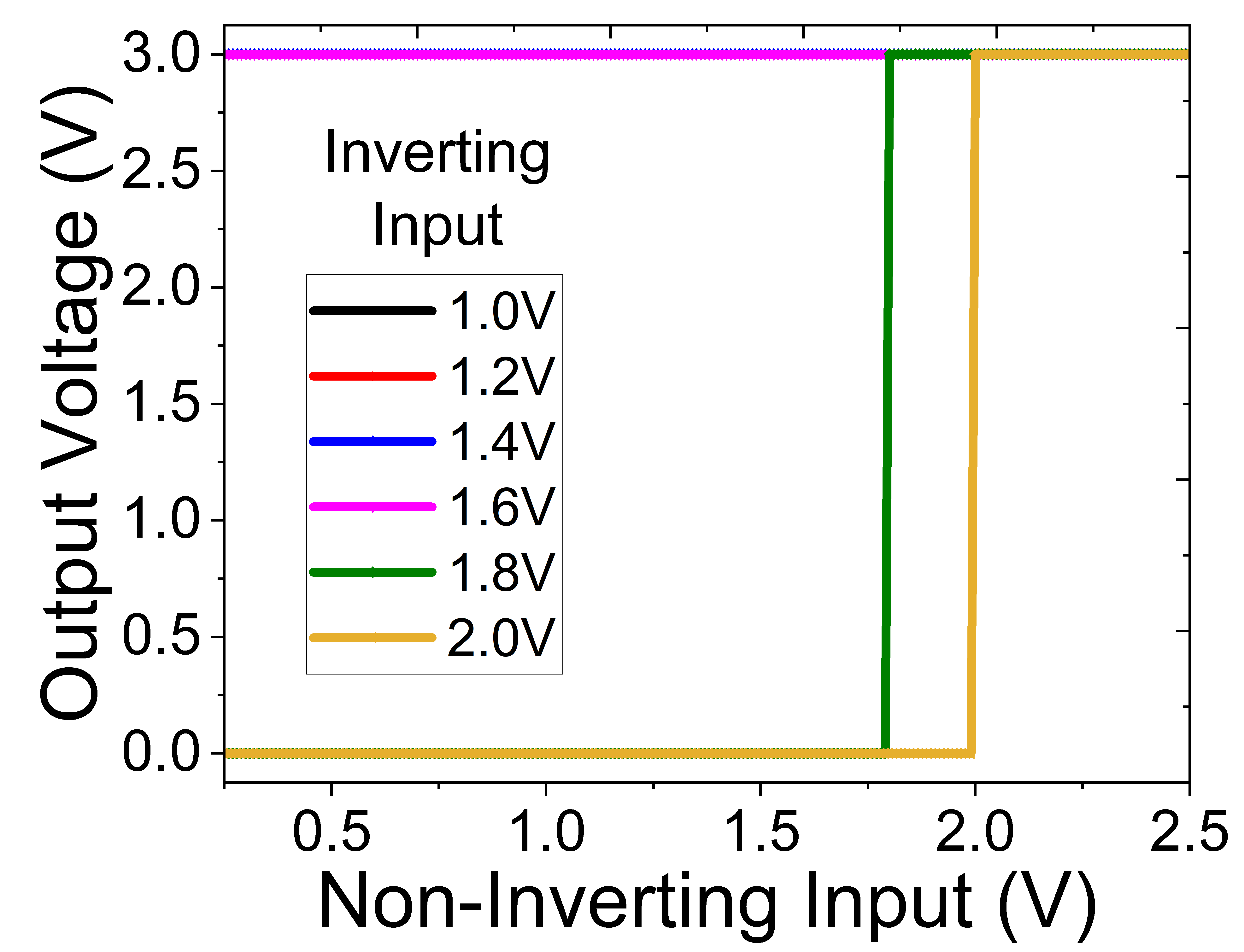}}%
\label{opamp_sens}
\hfil
\subfloat[]{\includegraphics[width=1.7in]{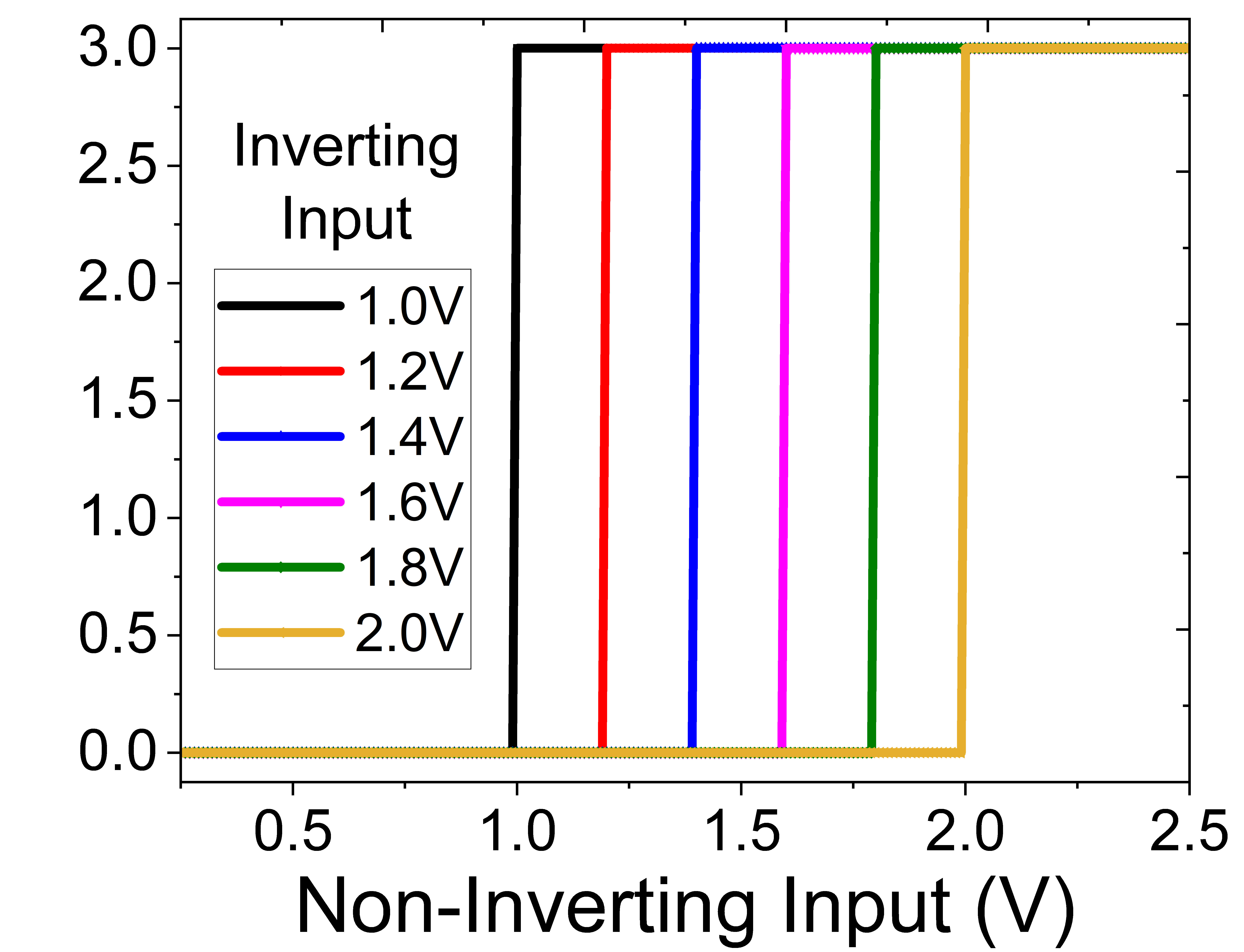}}%
\label{mp6541}
\caption{Plot of the non-inverting input voltage versus output voltage for the (a) TLV271I conventional operational amplifier and for the (b) MCP6541T-E/OT comparator at 77 K with supply voltage (VDD) of 3 V}
\label{sens}
\end{figure}

Measurements found that the in-operation of the previously proposed regulator is due to the OpAmp (Texas Instruments TLV271IDBVR) at cryogenic temperatures. At 77 K, the OpAmp's inputs exhibit a decreased sensitivity to lower voltages, requiring them to be within 1.2 V of OpAmp's power supply. This lack of sensitivity can be seen in Fig.~\ref{sens}(a). Ultimately, the need to supply multiple voltage domains spanning 0.6 V to 3.3 V with this constraint requires multiple unique pairs of voltage supplies and reference voltages. This practice ultimately diminishes the convenience that voltage regulators provide. For comparison, Fig.~\ref{sens}(b) shows the sensitivity of our proposed comparator that shows no signs of degraded sensitivity at 77 K.

%\begin{figure}[!h]
%\centering
%\subfloat[]{\includegraphics[width=1.5in]{Graphics/dilf_trim.png}%
%\hfil
%\subfloat[]{\includegraphics[width=1.5in]{Graphics/dilf_2.jpg}%

%\caption{(a) Mounted FPGA board with an SMA adaptor board located at the %4 K stage within the dilution fridge. (b) Both FPGAs installed in the %dilution fridge at two different temperature stages.}
%\label{dilf}
%\end{figure}

%Fig.~\ref{reg} displays the voltage regulation of the improved low-dropout regulator that is able to regulate with a single VDD supply by means of utilizing high-current comparators in place of the previously mentioned operational amplifiers (opamps) proposed by recent works. Fig.~\ref{opamp} gives reason for the previous work's non-operation as it displays the limited input sensitivity of general opamps when cold. In genera, it is required that the supply voltage (VDD) does not exceed 1.2 V above the intended regulated voltage. In this example, VDD is 3 V and thus cannot regulate voltages sweeps below 1.8 V as seen by the lack of transition to ground when the inverting input (VIN-) exceeds the non-inverting input (VIN+).

\begin{figure}[!h]
\centering
\subfloat[]{\includegraphics[width=1.7in]{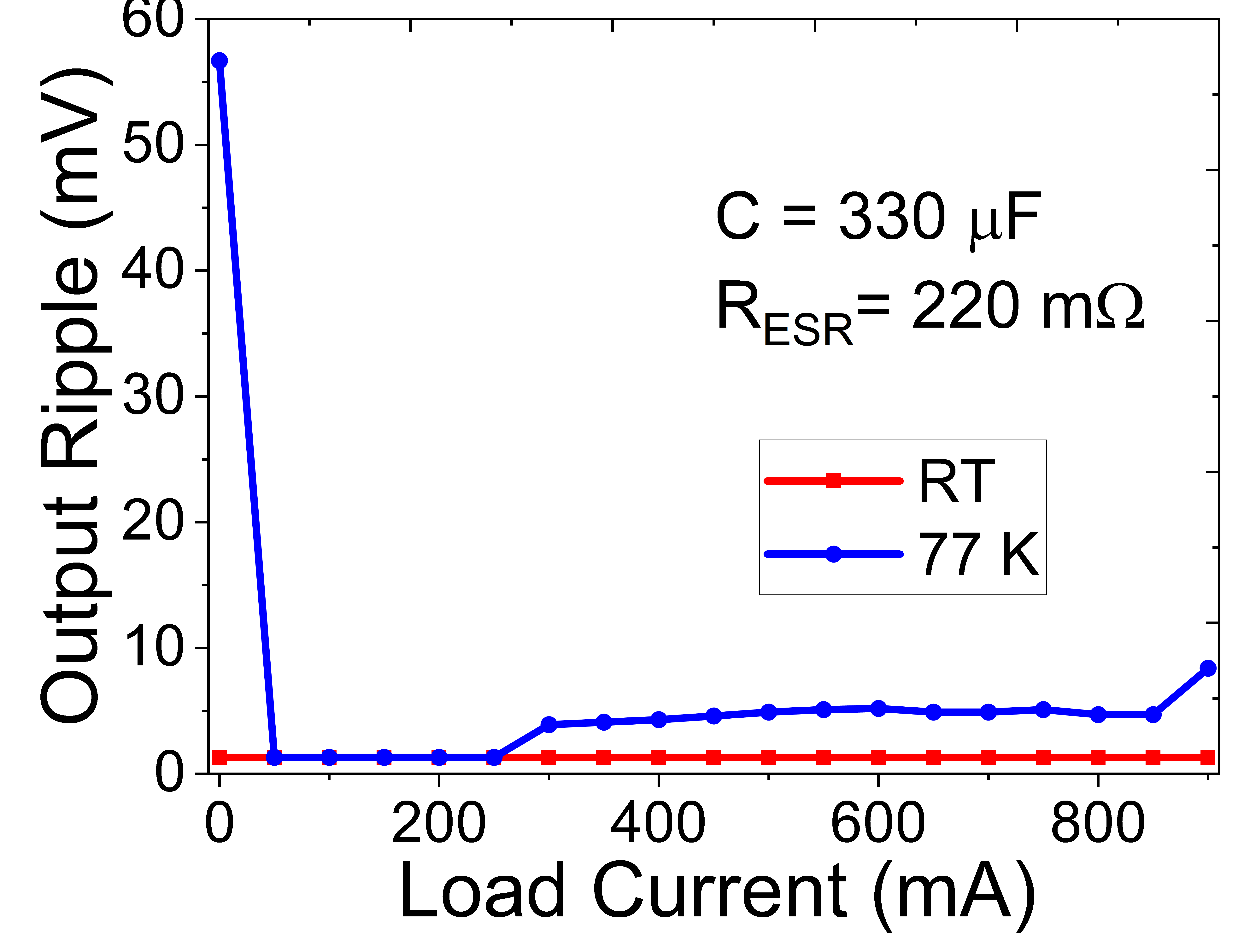}}%
\label{022_ripple}
\hfil
\subfloat[]{\includegraphics[width=1.7in]{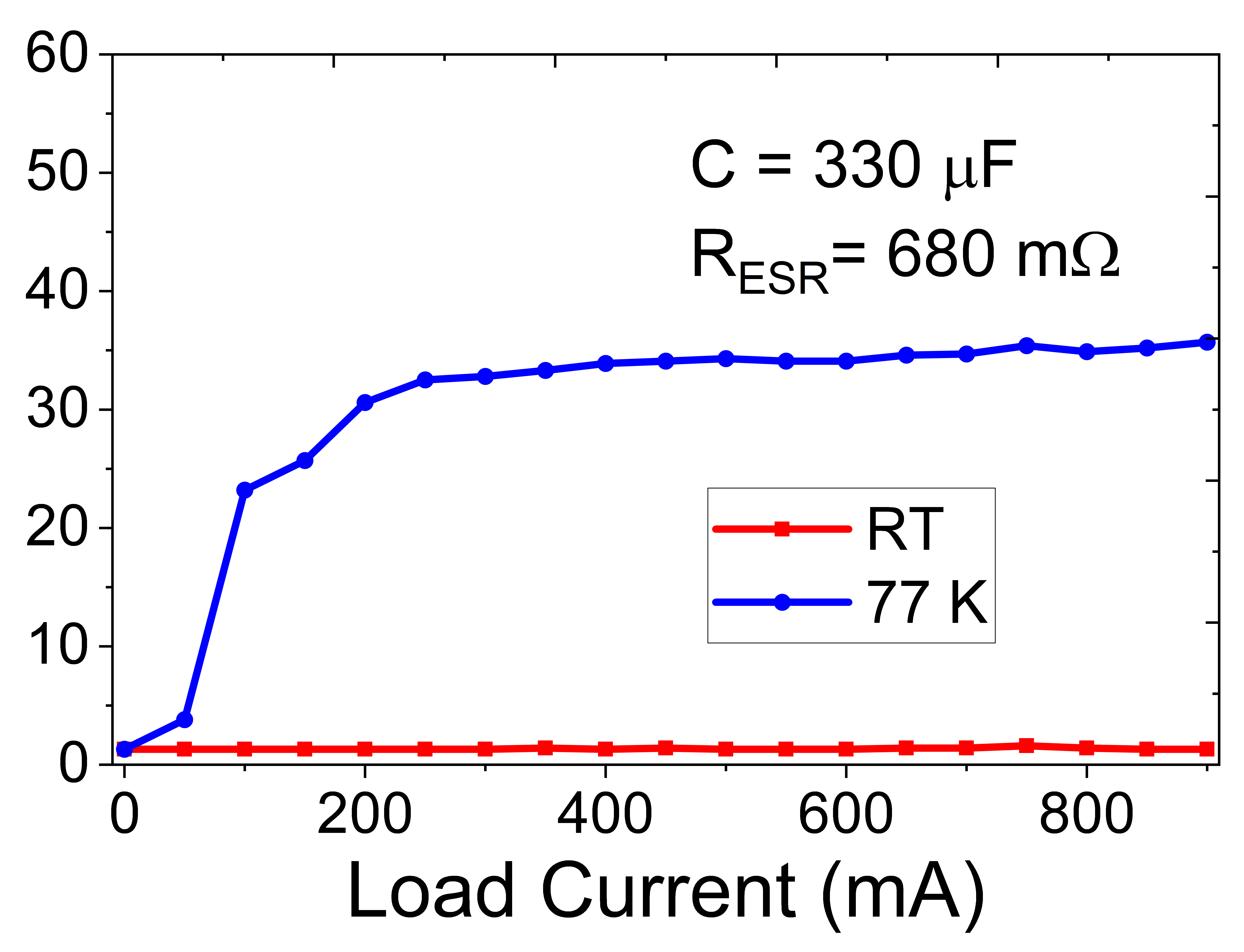}}%
\label{068_ripple}
\caption{Plot of our proposed LDO's output voltage ripple as a function of load current for an $R_{ESR}$ value of 220 $m\Omega$ (a) and 680 $m\Omega$ (b).}
\label{ripple}
\end{figure}

As mentioned previously, the use of this comparator leads to rippling at the regulator output requiring additional filtering at the comparator output, including the tuning of the capacitor's ESR via a metal-film resistor, $R_{ESR}$. The selection of this resistance became critical at lower temperatures and was observed to be load dependent, suggesting system instability. In experimentation, no single value for $R_{ESR}$ was found to be sufficient for all output load cases while 680 $m\Omega$ was determined to be effective for no load current and 220 $m\Omega$ was found to be effective for load currents greater than 50 mA. As a result, the output ripple voltage was reduced to less than 5 mV and as low as 1.5 mV across our load current range, 0 to 900 mA, as seen in Fig.~\ref{ripple}.

\begin{figure}[!h]
\centering
\subfloat[]{\includegraphics[width=1.7in]{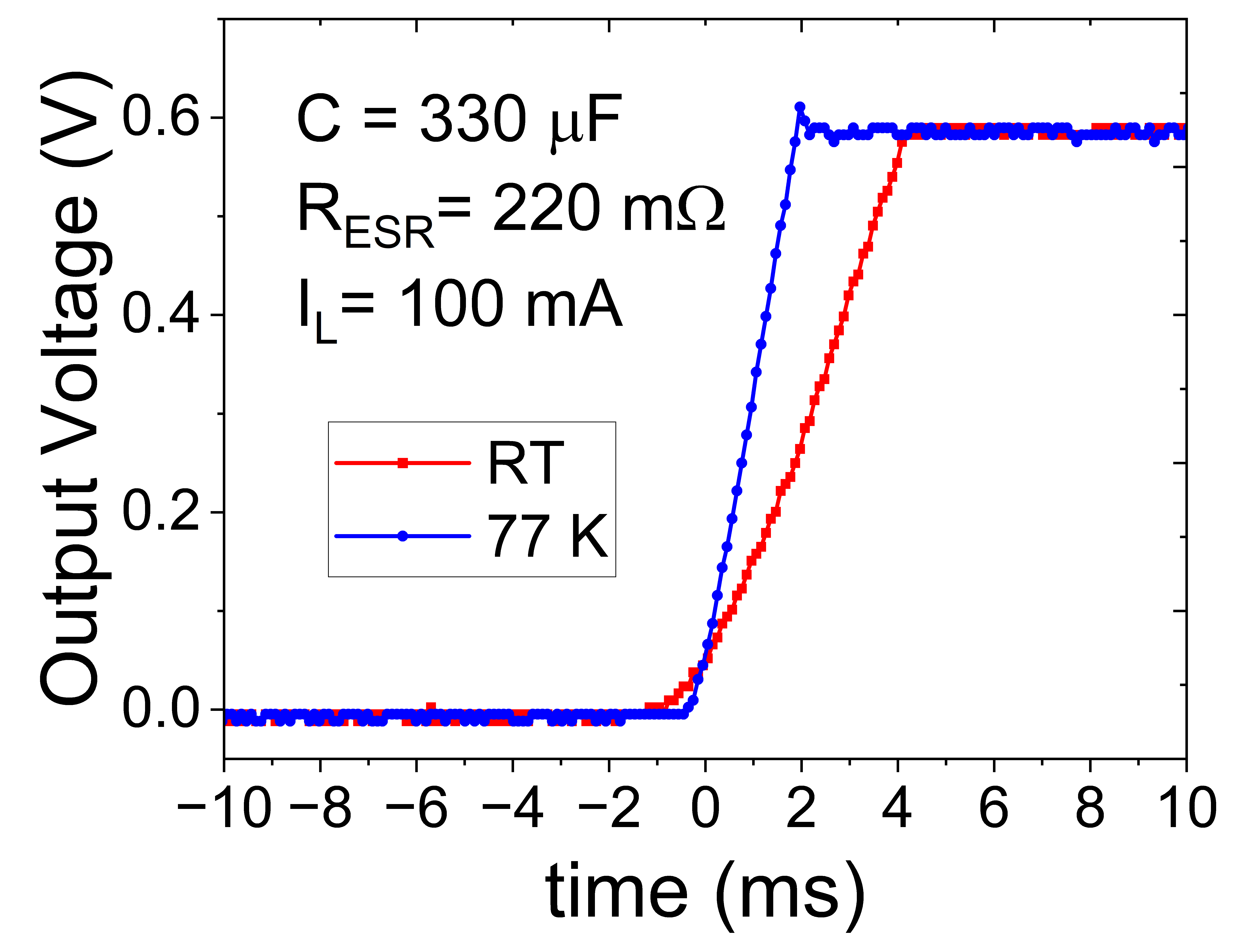}}%
\label{022_}
\hfil
\subfloat[]{\includegraphics[width=1.7in]{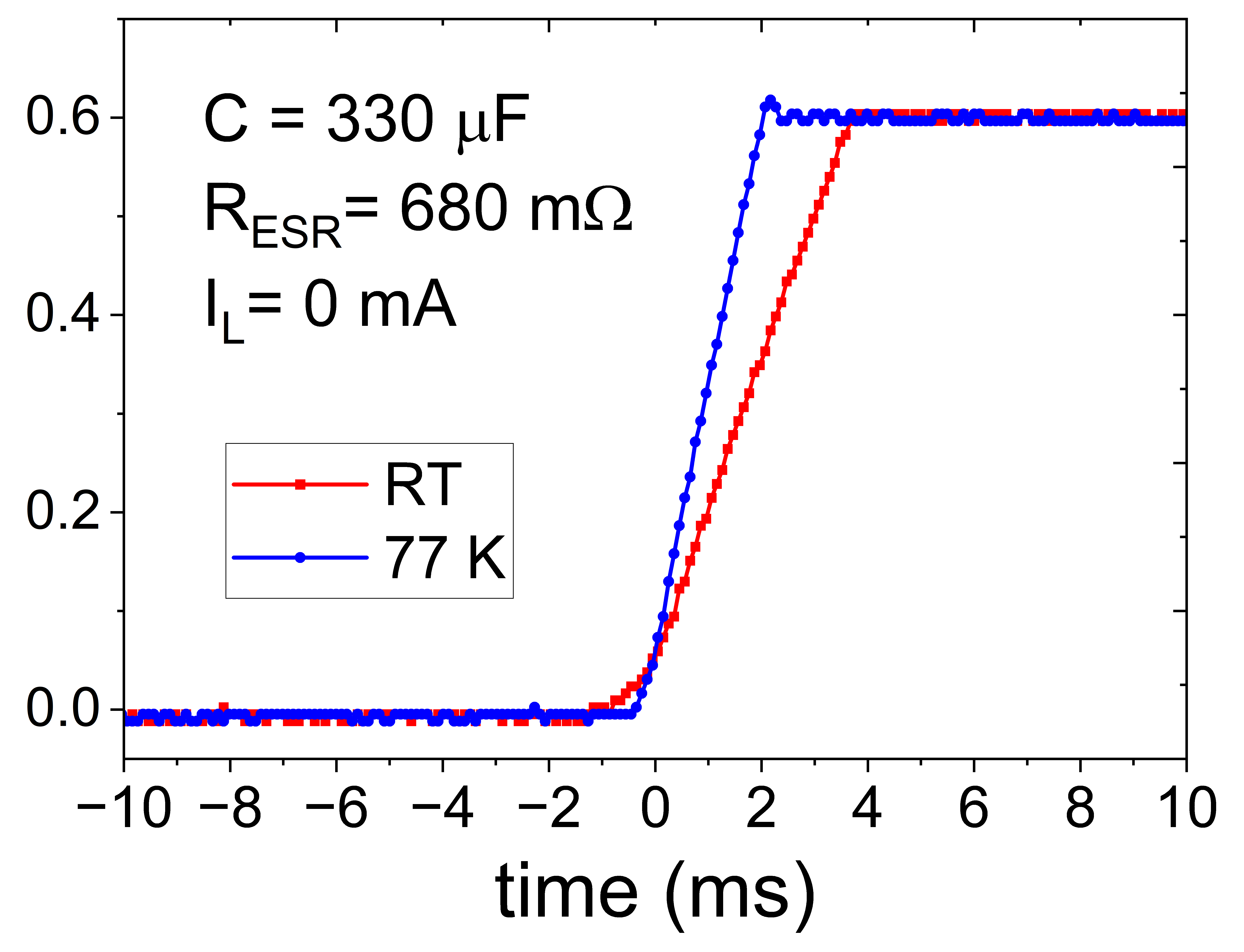}}%
\label{068_ripple}
\caption{Waveform of the LDO output voltage activating for a 600 mV target voltage for an $R_{ESR}$ value of 220 $m\Omega$ (a) and 680 $m\Omega$ (b) when coupled with a 235 uF load capacity.}
\label{on}
\end{figure}

\begin{figure}[!h]
\centering
\includegraphics[width=2.19in]{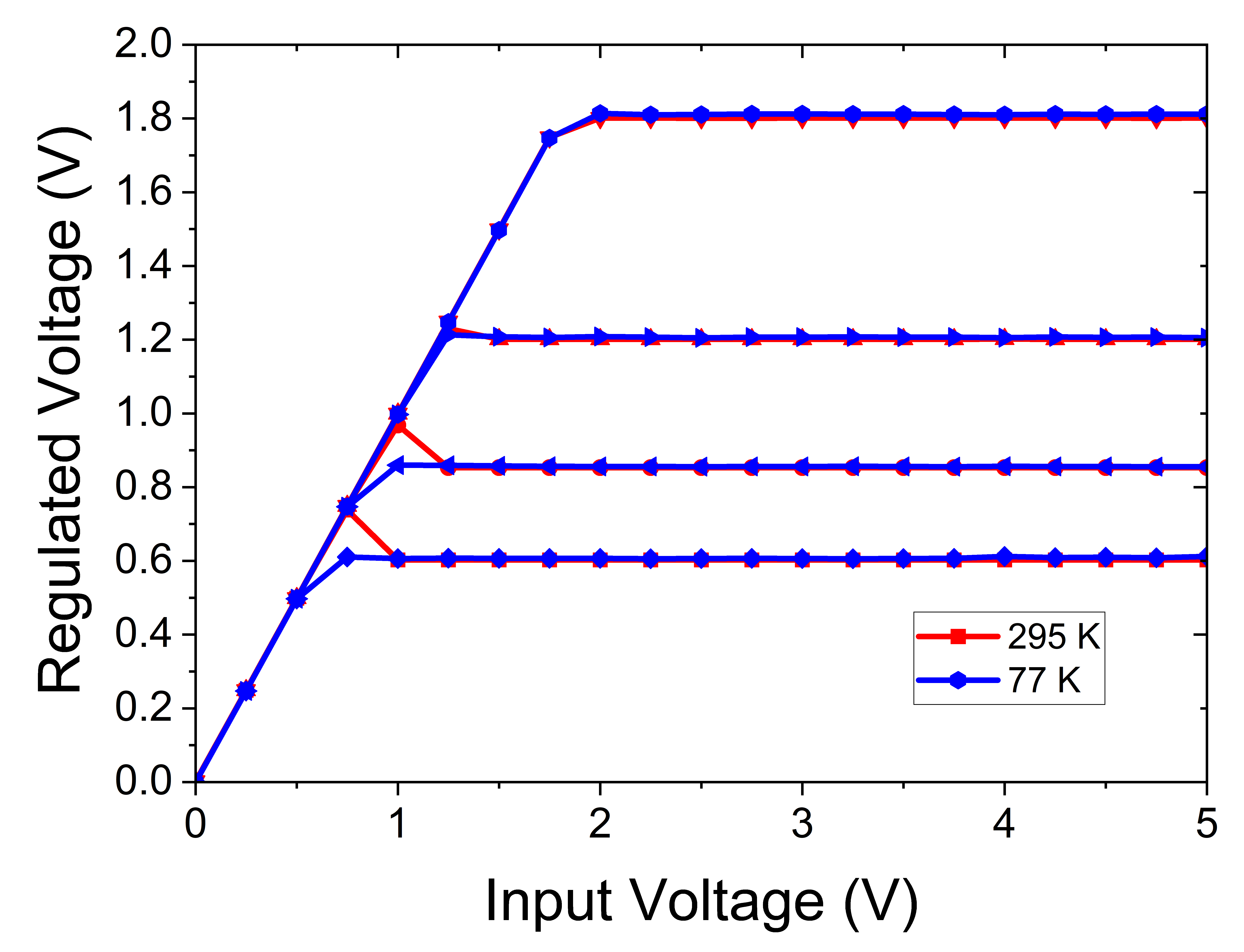}
\caption{Plot of input voltage versus the regulated output voltage for our proposed voltage regulator at various set voltages: 0.6 V, 0.85 V, 1.2 V, and 1.8 V for 77 K and room temperature (295 K).}
\label{reg}
\end{figure}

\begin{figure}[!h]
\centering
\includegraphics[width=2.19in]{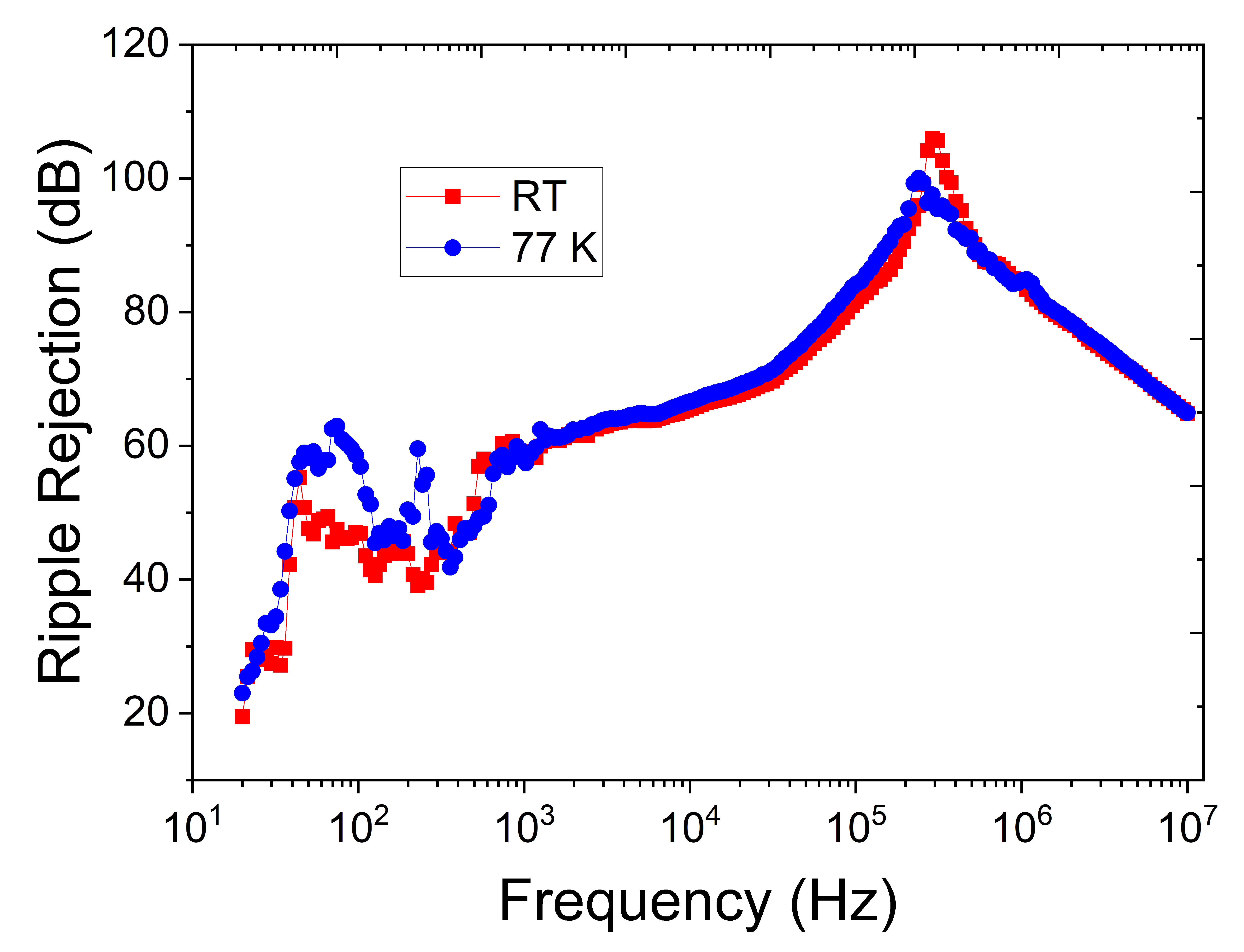}
\caption{Plot of LDO power supply ripple rejection (PSRR) ratio as a function of frequency for operation at 77 K and room temperature.}
\label{PSRR}
\end{figure}

\begin{figure}[!h]
\centering
\includegraphics[width=2.19in]{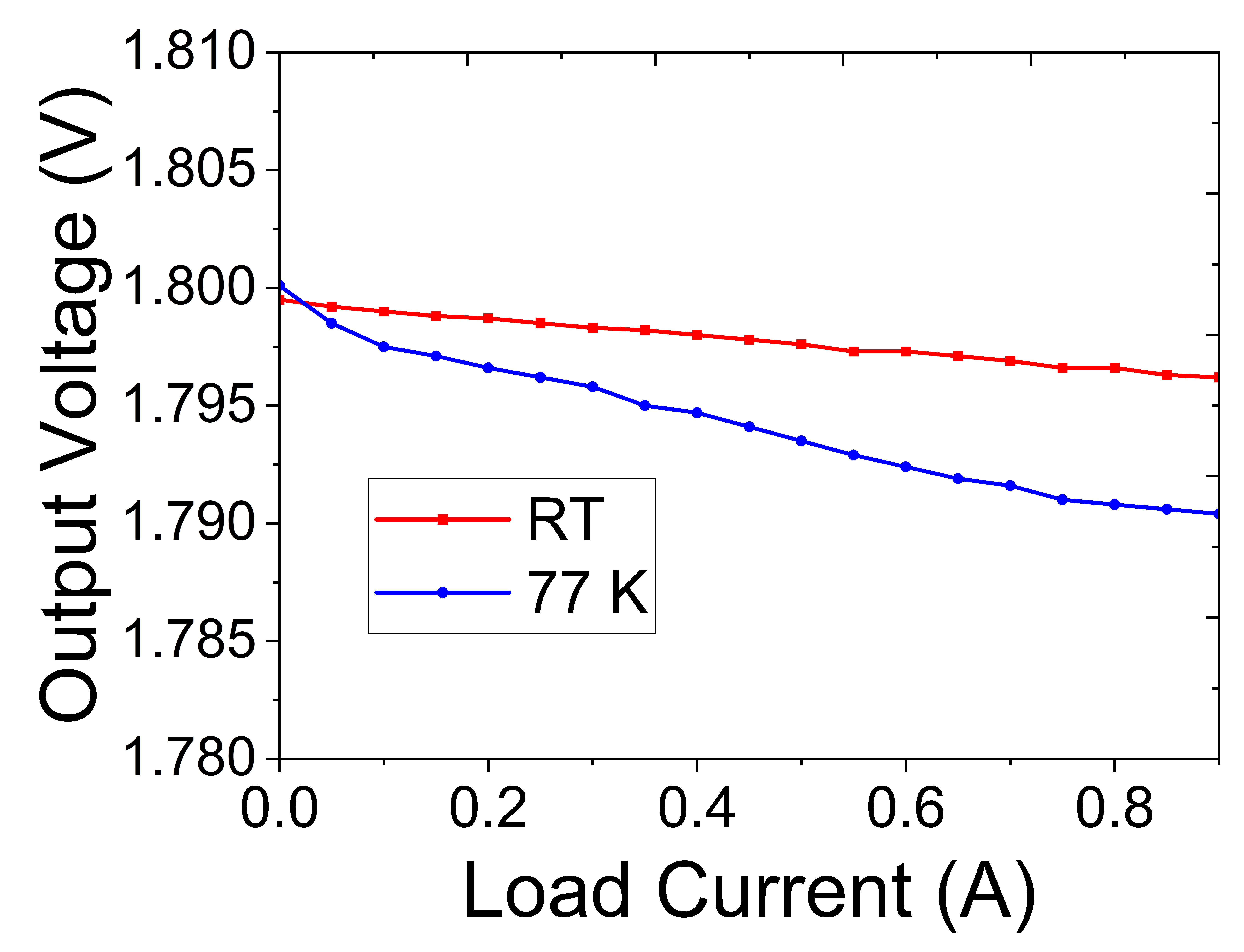}
\caption{Plot of output voltage as a function of load current at 77 K and room temperature with an input voltage of 2V.}
\label{load}
\end{figure}

The resulting performance of the LDO regulator is demonstrated through numerous metrics. Fig.~\ref{on} provides a waveform of the regulator's output when enabled demonstrating the regulator's turn-on rise time of approximately 4 ms at room temperature and seen to improve to just over 1 ms at 77 K for a 600 mV target voltage and a 235 uF load capacity, comparable to an FPGA power domain requirement. Little or no overshoot is observed, and an approximate slew rate of 539 mV/ms is observed and seen consistent for a target voltage of 1.8 V. For a zero-load current scenario, Fig.~\ref{reg} plots the regulator's output voltage versus its input voltage for various set voltages at 295 K and 77 K. This demonstrates the retained accuracy of the regulator even when cold and its ability to regulate despite an excessive input voltage. The regulator's power supply rejection ratio (PSRR) is also shown in Fig.~\ref{PSRR}, which was measured with an Agilent E5061B ENA network analyzer coupled with a bias tee for summing the AC and DC components of the LDO's input voltage for testing.  Lastly, the output voltage is plotted as a function of the load current in Fig.~\ref{load}. This displays a substantial amount of stability with large current loads displaying a loss of approximately 5 mV/A at room temperature that is seen to increase to approximately 10 mV/A for 77 K.

\section{Discussion}
The proof-of-concept design for the operation of commercial, high-performance 16-nm FinFET FPGAs at cryogenic temperatures is highly encouraging, as it demonstrates full operation despite operating far below its reported temperature range at 77 K while maintaining consistent power consumption. With adequate shielding for radiation hardening, these powerful FPGAs could see expanded usage in space applications spanning from satellites to astronomical space observatories given their wide temperature range towards cold, which can aid in reducing complex thermal requirements. 

Simultaneously, this is dramatic increase in power consumption compared to it's Artix-7 predecessors for the application of cryogenic computing. Given the limited cooling power of cryogenic refrigerators that can vary widely in several temperature stages, integration may be limited. In commercial cryostats, the available cooling powers typically range from 10 to 20 W for intermediate temperature stages ranging from 50 to 77 K. For 4 K and below, this cooling power is typically less than 600 mW. Given this information and the expectation for FPGA power consumption to increase at 4 K in tandem with it's predecessors, integration may be most suitable for intermediate temperatures stages as a likely candidate for control and readout of lower temperature electronics. Further power and design optimizations will be necessary for integration at lower temperatures, with there being attention most recently on the optimization at the device level to realize a true cryogenic CMOS FPGA \cite{optimization}.

The results of the design improvements for cryogenic low dropout regulators are also encouraging, as it presents a scalable solution for local power delivery for a fairly wide current range for cryogenic applications with it able to now power on from off while cold. As solutions for a stable output were found experimentally and vary by multiple factors such as load capacity and current, further modeling and stability analysis are warranted to realize usable models for effective circuit design, simulation, and control tuning.

\section{Conclusion}
This work has described and demonstrated the integration of commercial CMOS FPGAs utilizing the well-understood Xilinx Artix-7 FPGA as well as expanded to the unexplored Xilinx Zynq Ultrascale+ series of FPGAs for its expanded capabilities given its inclusion of gigabit-speed processors. To support these high-performance demands, improvements to existing work for cryogenic low dropout regulators were described, expanding their mode of operation. With this major step forward, high-performance CMOS FPGA technologies may see integration at cryogenic temperatures, suitable in areas of control and readout for cryogenic computing and satellite applications.

\section*{Acknowledgment}

The authors would like to thank the Alabama Micro/Nano Science and Technology Center (AMNSTC) for providing access to facilities used in this work and this work has been funded by the National Science Foundation (NSF) under the Expedition: DISCoVER (Design and Integration of Superconducting Computation for Ventures beyond Exascale Realization) project grant number 2124453.

%%%%%%%%%%%%%%%%%%%%%%%%%%%%%%%%%%%%%%%%%%%%%%%%%%%%%%%%%%%%%%%%%%%%%%%%%%%%%%%%

%References are important to the reader; therefore, each citation must
%be complete and correct. If at all possible, references should be
%commonly available publications.

\bibliographystyle{IEEEtran}
\end{document}